\documentclass{article}
\usepackage{arxiv}
\usepackage{orcidlink}
\usepackage{hyperref}

\providecommand{\orcidlinkf}[1]{%
\orcidlink{#1}\,\url{https://orcid.org/#1}%
}

\renewcommand{\headeright}{}
\renewcommand{\undertitle}{}

\usepackage{paralist}

\usepackage{mathtools}

\newcommand{\ceil}[1]{\lceil {#1} \rceil}

\usepackage{enumitem}

\usepackage{parskip}

\usepackage{helvet}

\usepackage{caption}
\usepackage{subcaption}

\usepackage{graphicx}

\newcommand{\Connected}{\mathsf{Connected}}

\newcommand{\MaxCycle}{\mathrm{MaxCycle}}
\newcommand{\NumStages}{\mathrm{NumStages}}
\newcommand{\MaxNumStages}{\mathrm{MaxNumStages}}

\newcommand{\Count}{\mathrm{count}}
\newcommand{\Cycle}{\mathrm{Cycle}}
\newcommand{\Slot}{\mathrm{Slot}}
\newcommand{\Start}{\mathrm{Start}}
\newcommand{\End}{\mathrm{End}}
\newcommand{\II}{\mathrm{II}}
\newcommand{\MaxUse}{\mathrm{MaxUse}}
\newcommand{\MinUse}{\mathrm{MinUse}}
\newcommand{\MaxUseBeforeDef}{\mathrm{MaxUseBeforeDef}}

\newcommand{\Used}{\mathrm{Used}}
\newcommand{\UsedBeforeDef}{\mathrm{UsedBeforeDef}}
\newcommand{\LiveOut}{\mathrm{LiveOut}}
\newcommand{\Live}{\mathrm{Live}}

\newcommand{\weg}[1]{}

\newcommand{\TrajB}[1]{\bt{\TrajB}(#1)}

\newcommand{\bt}[1]{#1^{\leftrightarrows}}

\title{Optimal Software Pipelining using an SMT-Solver}
\author{
Jan-Willem Roorda\thanks{\orcidlinkf{0009-0003-9218-7035}}\\
Intel Corporation\\
}

\date{}

\begin{document}

\maketitle

\begin{abstract}
Software Pipelining is a classic and important loop-optimization for VLIW processors. It improves instruction-level parallelism by overlapping multiple iterations of a loop and executing them in parallel. Typically, it is implemented using heuristics. In this paper, we present an optimal software pipeliner based on a Satisfiability Modulo Theories (SMT) Solver. We show that our approach significantly outperforms heuristic algorithms and hand-optimization. Furthermore, we show how the solver can be used to give feedback to programmers and processor designers on why a software pipelined schedule of a certain initiation interval is not feasible

\end{abstract}	

\section{Introduction}

\subsection{VLIW}
The \emph{Very Long Instruction Word (VLIW) architecture}\cite{VLIWBook} refers to a
class of processors that
\begin{enumerate}
\item contain multiple issue slots such that operations can be executed in parallel, and
\item expose that parallelism directly to the compiler.
\end{enumerate}

In contrast to \emph{super-scalar} processors, which contain hardware that at
run-time decides which operations can be scheduled in parallel,
VLIW-processors leave this task completely to the compiler.
That is, the compiler decides for each cycle and issue slot, which machine
operations are executed, in which registers the results of the operations are
stored, and how these results are routed to the register-files via buses and
register-file write-ports.

Letting the compiler at compile time 
(instead of the hardware at run time) decide
which operations can be scheduled
makes it possible to omit hardware blocks dedicated to run-time scheduling, such as
reorder buffers and register-renaming units.
This can save considerable amounts of area and power.
This makes VLIW-processors an ideal
candidate for applications that
have tight power and area constraints but also require the flexibility of
a programmable solution.

\subsection{Compiling for VLIW}
Compiling for VLIW is typically more challenging than compiling for super-scalar processors.
There are several reasons for this:
\begin{itemize}
\item{All schedule decisions have to be taken by the compiler at compile time. The compiler cannot, as with super scalar architectures,
rely on the processor to take part of the scheduling burden.}
\item{VLIW cores typically contain a high number of issue-slots. To efficiently use the core, the compiler should make sure that the
generated schedule contains enough instruction level parallelism.}
\item{The generated schedule directly determines the performance of the compiled application. For instance, scheduling an inner loop in 10 cycles
  instead of 9, gives a performance degradation of that inner loop of a factor 0.9. }
\end{itemize}
Optimization is typically focused on the inner loops of the code. Unfortunately, in many
cases those inner loops do not allow enough parallelism to keep
all the issue slots of a VLIW-processor busy.
Two common code transformations that can be used in such cases are
\emph{Loop Unrolling} and \emph{Software pipelining (SWP)}. We will explain both below.

\subsection{Loop Unrolling}

\begin{figure}
\begin{subfigure}{.5\textwidth}
  \centering
  \begin{verbatim}
int A[1024] // input array
int B[1024] // output array
for(int i=0; i<1024; i++) {
  a  = LD A
  b  = a+2
  c  = b*2
  ST c
}
  \end{verbatim}
  \caption{Example Program. We assume the load (LD) and store (ST) instructions are auto-incrementing. For simplicity, we have left out the offset variables for the load and store. }
  \label{fig:program}
\end{subfigure}%
\begin{subfigure}{.5\textwidth}
  \centering
  \includegraphics[width=.7\linewidth]{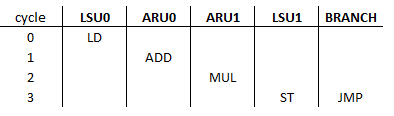}
  \caption{A schedule for the loop from the program on the left.}
  \label{fig:schedule}
\end{subfigure}
\caption{}
\label{fig:test}
\end{figure}

As an example, consider the pseudo-code in Figure \ref{fig:program}. Let us assume that the program will be scheduled on a VLIW-processor
with five issue-slots consisting of two arithmetic slots (ARU0 and ARU1), two load/store slots (LSU0 and LSU1), and a branch unit (BRA).
For simplicity, we assume that each instruction takes 1 clock cycle and there is no
branch delay.

Without applying any loop transformations, there is not much parallelism in the schedule for this code:
At each cycle only one of the four non-branch issue-slots is used. See figure \ref{fig:schedule}.
Executing the loop will take $1024 \times 4 = 4096$ cycles.

\begin{figure}
\centering
\begin{subfigure}{.5\textwidth}
  \centering
  \includegraphics[width=.7\linewidth]{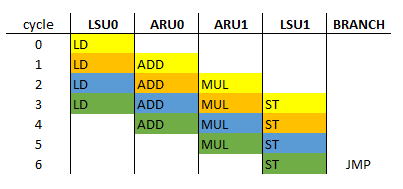}
  \caption{The schedule for the loop in the program of Figure \ref{fig:program} unrolled 4 times.}
  \label{fig:schedule_unroll4}
\end{subfigure}%
\begin{subfigure}{.5\textwidth}
  \centering
  \includegraphics[width=.7\linewidth]{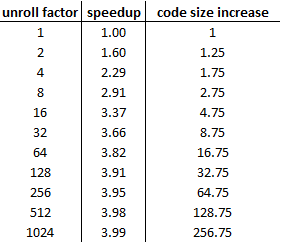}
  \caption{The effect of unrolling on performance and code size.}
  \label{fig:unroll_effects}
\end{subfigure}
\caption{}
\end{figure}

A straightforward way of optimizing this loop is to apply loop unrolling.
Unrolling the loop four times yields the schedule in Figure \ref{fig:schedule_unroll4}.
As illustrated, when loop unrolling, the body of a loop is replicated multiple times. In this case, the body of the newly created loop contains instructions from four iterations (in the picture, each iteration is given a different color).
This creates more opportunity for parallelism, improving the efficiency of the execution. 
Executing the unrolled loop takes $256 \times 7 = 1792$ cycles, a speed-up of a factor 2.29. The drawback of unrolling is the increased code size. 
In this case, the code size has increased by a factor $1.75$ to $7$ cycles.

So, we see that increasing unrolling leads to increased performance, but, unfortunately, also to an increased code-size.
The more we unroll, the higher the performance, but also the larger the code size.
The table in Figure \ref{fig:unroll_effects} illustrates this for this example.

\subsection{Software Pipelining}

\begin{figure}
\centering
\begin{subfigure}{.5\textwidth}
  \centering
  \includegraphics[width=.7\linewidth]{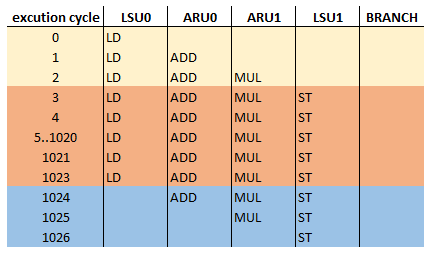}
  \caption{The schedule for the full unroll of the loop in Figure \ref{fig:program}. 
}
  \label{fig:trace_full_unroll}
\end{subfigure}%
\begin{subfigure}{.5\textwidth}
  \centering
  \includegraphics[width=.7\linewidth]{schedule_unroll4.png}
  \caption{The schedule of the software pipelined loop in the program of Figure \ref{fig:program}}
  \label{fig:swp}
\end{subfigure}
\caption{}
\end{figure}

The intuition behind Software Pipelining can be explained by looking at the schedule of the full unroll of the loop in our example in
Figure \ref{fig:trace_full_unroll}. 

In the schedule, we can recognize a pipeline: 
the first three cycles (in yellow) are used to fill the pipeline. Then in cycles 3 until 1023 (in red), the pipeline 
executes at full speed. And, finally, the last three cycles (in blue) are used to drain the pipeline.

The idea of Software pipelining is that given a loop, three blocks should be created: the \emph{prolog} filling the pipeline.
The \emph{kernel} (a single basic block loop) starts a new iteration of the original loop each time it is executed.
And, finally, the \emph{epilog} that drains the pipeline.

For our running example, the software pipelined schedule is displayed in Figure \ref{fig:swp}. 
When it is executed it has the same performance as the fully unrolled 
loop, but it only takes a fraction of the code-size.

Software Pipelining allows to approach, and sometimes even reach, 
the performance of full unrolling with a limited increase in code-size. 
For this reason, it is an essential component of an optimizing compiler
for VLIW processors.

\subsection{Implementations of Software pipelining}

Although for our running example finding a software pipelined schedule is very easy, the general SWP-problem is NP-complete.
For that reason compilers typically use \emph{heuristic} methods to implement SWP. Two popular heuristic algorithms
are \emph{Iterative Modulo Scheduling (IMS)} \cite{ModSchedRau,IMSRau} and \emph{Swing Modulo Scheduling}\cite{SMS1,SMS2}.

Next to heuristics, also \emph{optimal methods} for SWP based on \emph{Integer Programming (IP)}\cite{IPBook}
have been described. For an overview, see \cite{ExhSchedSurvey}.

\subsection{SAT and SMT}
A \emph{Boolean Satisfiability (SAT) Solver} decides for a formula in propositional logic
whether it is satisfiable\cite{grasp}. That is, it decides whether there is
an assignment to the variables under which the formula is true.

In recent years, the power of satisfiability solvers, often called SAT-solvers,
has increased dramatically. SAT-solvers have successfully been used in domains
varying from hard- and software verification, automatic test-pattern
generation, planning and scheduling, to bio-informatics\cite{practicalsat}.

\emph{Satisfiability modulo theories}\cite{SMTBook} (SMT) generalizes the Boolean satisfiability
problem to constraints including integers, real numbers, quantifiers, arrays, etc.

\subsection{Our contribution} 

In this paper, we present an optimal software pipelining algorithm based on a Satisfiability Modulo Theories (SMT) Solver.
To our knowledge, we are the first to do so.

Our algorithm is used to schedule all (400+) software pipelined loops in the firmware code for two generations of VLIW-processors used in
signal processing for wireless communication.

We give experimental results in which we show that our method compares favorably with heuristic algorithms and hand-optimization:
On our benchmark set, the maximum measured speed-up was $1.22$. The geometric average speed-up was $1.08$. 
See Section \ref{sec:results} for more details.

We are the first to show how the \emph{unsatisfiable core} generation feature of modern SMT-solvers can be used by the compiler
to give feedback to programmers and processor-designers about \emph{why} a software pipelined schedule for a certain II is not possible.

\subsection{Example of the use of the UNSAT-core}

When finding that an SMT-problem is not satisfiable, the SMT-solver can output
an \emph{Unsatisfiable Core}. This is a subset of the set of constraints
that is unsatisfiable by itself.

By printing the constraints in the unsatisfiable core in a 
human readable format, the compiler can explain \emph{why} 
a software pipelined schedule of a certain length is unfeasible.

\begin{figure}
 \centering
  \includegraphics[width=.7\linewidth]{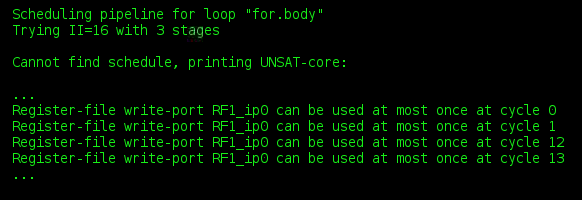}
  \caption{An example of an unsatisfiable core. }
  \label{fig:unsat}
\end{figure}

For instance, in Figure \ref{fig:unsat} a part of the unsatisfiable 
core is printed. From this, the user can infer that a reason for
not being able to find a software pipelined schedule of length 16 is that
write port \textrm{ip0} of register file \textrm{RF1} is overloaded. A solution
for this could be to change the processor design and/or the firmware 
such that the write port is used less often.

\section{Description}

\subsection{Modulo Scheduling}

The intuition behind the \emph{Modulo scheduling framework} is as follows\cite{IMSRau}: 
Suppose that we unroll a loop completely, and then schedule the code with two constraints:
\begin{itemize}
\item{all iterations have identical schedules, except}
\item{each iteration is scheduled a fixed number of cycles (called the \emph{Initiation interval} (II)) after the previous iteration.}
\end{itemize}

A schedule meeting the above constraints consists of
(1) a part filling the pipeline, (2) a large repetitive part, and (3) a part draining the pipeline.
The idea is that the repetitive part can compactly be represented by a kernel loop.
The part of the schedule filling the pipeline is called the prolog, the part that drains it
is called the epilog.

In  modulo scheduling, instead of fully
unrolling the loop, a schedule is constructed for a \emph{single} iteration of the loop such that 
when the same schedule is repeated at intervals of II cycles the following holds:
\begin{enumerate}
\item{No resource conflict arises between different iterations}
  
To do so, we need to make sure that in the schedule no resources
are used both at cycle $c_1$ and $c_2$ when $c_1 \% \mathrm{II}=c_2 \% \II$. This is where
the name \emph{modulo} scheduling comes from.
\item{No (loop-carried) dependencies are violated.}
 
Next to usual dependencies, like data-, anti-, and output dependencies, we also
need to take into account \emph{loop-carried dependencies} that express
scheduling constraints between operations of different iterations of the loop.

For instance, suppose we have an operation $a$ that writes a value 
read by operation $b$. This causes a (normal, not loop-carried)
dependency with a latency of 1 between operation $a$ and $b$. 

However, when modulo scheduling, we also need to make sure that operation $b$ reads 
the value written by operation $a$ of the \emph{current} iteration \emph{before} operation $a$ of the 
\emph{next} iteration overwrites it with a potentially different value.
To do so, we introduce a \emph{loop-carried dependency} between $b$ and $a$
with latency $0$ and \emph{distance} $1$. Here, the distance indicates the number of
iterations $b$ and $a$ are apart. So, the dependency means
that operation $b$ from iteration $n$ should be scheduled before operation $a$
of iteration $n+1$.
In the modulo schedule for a specific $\II$, each $\II$ cycles a new iteration
is started. So, to meet this dependency constraint $b$ should be scheduled at most $\II$ cycles
after $a$.
In general, to meet a dependency constraint between $o_1$ and $o_2$ with latency $l$
and distance $d$, the following should hold\cite{IMSRau}:
\[
\Cycle(o_2) \geq \Cycle(o_1) + l - d \times \II   
\]
\end{enumerate}

When a schedule for such a single iteration can be found, a software pipelined schedule
consisting of a prolog, kernel loop, and epilog can be constructed from it.
For instance, the schedule for the kernel of the SWPed loop can be created
by mapping each operation scheduled at cycle $c$ of the modulo schedule to
cycle $c\%\II$ in the kernel. 

The number of overlapping iterations in the SWPed schedule is called the number of \emph{stages}.
It can be calculated by dividing the length of the schedule by II and rounding it up.

Both \emph{Iterative Modulo Scheduling} \cite{ModSchedRau,IMSRau} 
and \emph{Swing Modulo Scheduling}\cite{SMS1,SMS2} are heuristic algorithms 
based on the Modulo Scheduling framework. 
In the remainder of this section we explain how we perform optimal Modulo Scheduling using an SMT-Solver.

\subsection{the SMT Problem for SWP}
\label{sec:smt-problem}
Below, we explain how, for a given II and number of stages, we can use an SMT-solver to find 
valid modulo schedules.
For now, we will not consider register allocation. We explain how we deal with register allocation in Section \ref{sec:rp}.

Let $\MaxCycle$ be the last cycle in the schedule for a single iteration. 
It is defined by: $\MaxCycle= \II  \times \NumStages-1$.

We use the following variables in the SMT-problem:

\begin{enumerate}
\item{For each operation $o$, an integer variable $\Cycle_o$ modeling the cycle on which it is scheduled}
\item{For each operation $o$ and issue slot $s$, a Boolean variable $\Slot_{o,s}$ modeling whether the operation is scheduled on the issue-slot}
\item{For each cycle $c$ with $0 \leq c \leq \MaxCycle$, issue-slot output port $op$, and bus $b$, 
a Boolean variable $\Connected_{c,op,b}$ modeling whether at cycle $c$ the connection
between issue-slot output-port $op$ and bus $b$ is made}. 
\item{For each cycle $c$ with $0 \leq c \leq \MaxCycle$, bus $b$, and register-file write port $wp$,
 a Boolean variable $\Connected_{c,b,wp}$ modeling whether at cycle $c$ the connection
between bus $b$ and register-file write-port $wp$ is made}. 
\end{enumerate}

Each assignment to the variables corresponds to a modulo schedule. The constraints below make sure that
we only consider \emph{feasible} modulo schedules.

\begin{enumerate}
\item{Each operation is scheduled on a valid cycle.}

For each operation $o$: 
\[
0 \leq \Cycle_{o} \leq \MaxCycle   
\]

\item{Each operation is scheduled on exactly one issue-slot.}

For each operation $o$ and set of issue-slots $S$ that $o$ can be scheduled on:
\[ \bigoplus\limits_{s \in S} \Slot_{o,s} \]

Here $\bigoplus$ denotes the $n$-ary xor function, that returns true when
exactly one of its arguments is true. 
\item{Each issue-slot is used only once per modulo cycle.}

For all cycles $c_1$ and $c_2$ such that $c_1\%\II=c_2\%\II$, slot $s$, and distinct operations $o_1$ and $o_2$: 
\[
  \neg(\Slot_{o_1,s} \land \Slot_{o_2,s} \land \Cycle_{o_1}=c_1 \land \Cycle_{o_2}=c_2)
\]

\item{All dependencies are met.}

For each dependency between operations $o_1$ and  $o_2$ with latency $l$ and distance $d$:
\[
\Cycle_{o_2} \geq \Cycle_{o_1} + l - d \times \II   
\]

\item{For each data-flow dependency, there is a connection to route the output of the issue-slot to the correct register-file}

For each data flow dependency between
 operations 
 $o_1$ and 
 $o_2$, 
 issue-slots $s_1$ and $s_2$, cycle $c$, 
 let $op$ be the issue-slot outport that $o_1$ writes the data to 
 when it is scheduled on slot $s_1$,
 let $W$ be the set of register-file write-ports of the register-file that
 $o_2$ reads the data from when it is scheduled on $s_2$, 
 and
let $\mathrm{B}(wp,op)$ be the set of buses that connect 
register-file write-port $wp$ with issue-slot output-port $op$
\[
  \Slot_{o_1,s_1} \land \Slot_{o_2,s_2} \land \Cycle_{o_1}=c  \implies \bigvee\limits_{\substack{wp\in W \\ b \in \mathrm{B}(wp,op)}} (\Connected_{c,op,b} \land \Connected_{c,b,wp})
\]

Here $\bigvee$ denotes the $n$-ary or function, that returns true when
at least one of its arguments is true. 

\item{Each bus is connected to at most one issue-slot output port per modulo cycle.}

For all cycles $c_1$ and $c_2$ such that $c_1\%\II=c_2\%\II$, bus $b$,
 and distinct issue-slot output ports $op_1$ and $op_2$: 
\[ \neg( \Connected_{c_1,b,op_1} \land \Connected_{c_2,b,op_2}) \]

\item{Each register-file write-port is connected to at most one bus per modulo cycle}

For all cycles $c_1$ and $c_2$ such that $c_1\%\II=c_2\%\II$,
register-file write port $wp$, and distinct buses $b_1$ and $b_2$: 
\[ \neg( \Connected_{c_1,b_1,wp} \land \Connected_{c_2,b_2,wp}) \]

\end{enumerate}

\subsection{The exhaustive algorithm}

\begin{figure}
\centering
\begin{subfigure}{.5\textwidth}
  \centering
  \includegraphics[width=.7\linewidth]{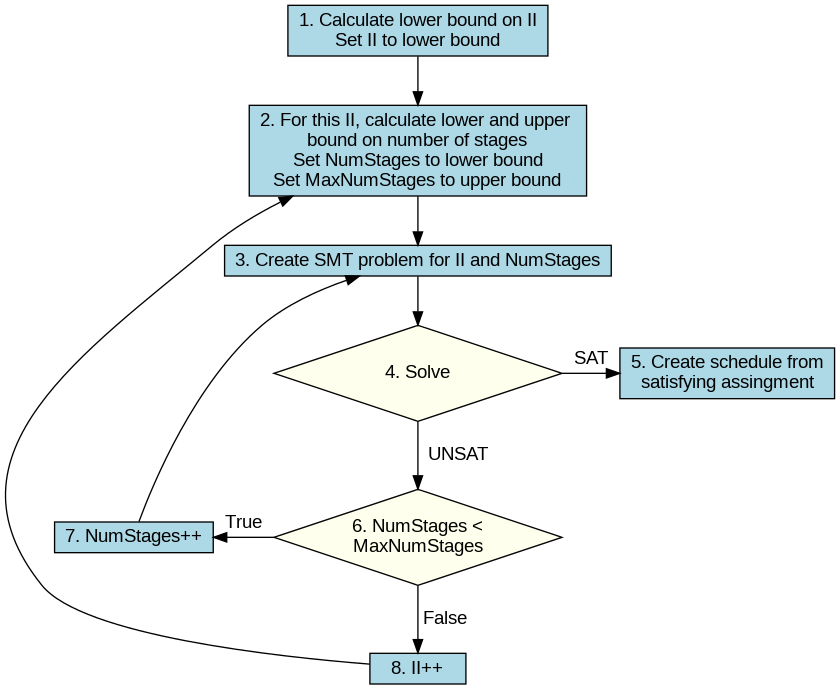}
  \caption{The algorithm without taking register pressure\\ into account}
  \label{fig:algo}
\end{subfigure}%
\begin{subfigure}{.5\textwidth}
  \centering
  \includegraphics[width=.7\linewidth]{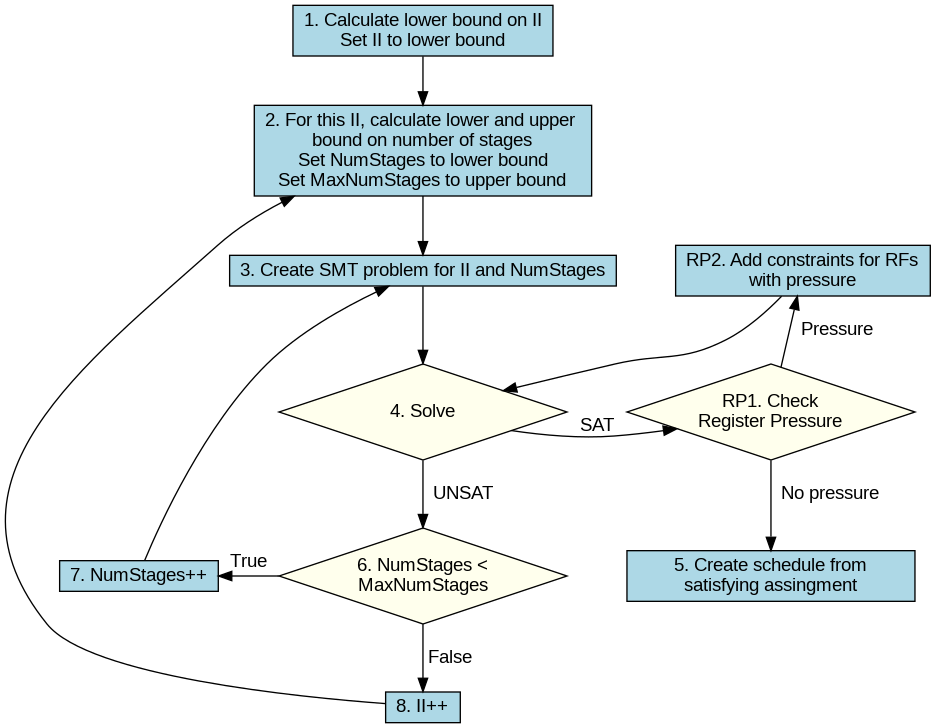}
  \caption{The algorithm with register pressure taken into account}
  \label{fig:algo_rp}
\end{subfigure}
\caption{}
\end{figure}

The exhaustive algorithm we use for finding a software pipelined schedule with a minimal $\II$ is given in Figure \ref{fig:algo}. 
Below we explain the algorithm in more detail:

In \textbf{step 1}, we determine a lower bound for the initiation interval of the software pipelined loop.
How to find this lower bound is well known\cite{IMSRau} and, therefore, we will not describe it here.
We set the variable $\II$
to this lower bound.

In \textbf{step 2}, we calculate (from the given $\II$) a lower bound and an upper bound on the number of stages in the modulo schedule. 
The lower bound is determined by dividing the critical path of the dependency graph by the $\II$.
We set $\NumStages$ to this lower bound.

The intuition of the upper bound is based on the fact that the by SWP induced loop-carried dependencies
impose a \emph{maximum}  number of cycles that operations may be apart. 
For instance, if there is a loop carried dependency with latency $l$ and distance $d$ 
between operations $o_1$ and $o_2$ then the following must hold $\Cycle(o_2) \geq \Cycle(o_1) + l - d \times \II$.
This is equivalent to $  \Cycle(o_1) \leq \Cycle(o_2) + d \times \II - l$. This means that operation $o_1$ 
may be scheduled \emph{at most} $d \times \II - l $ cycles later than operation $o_2$.

As described above, for each data-flow edge between $a$ and $b$ in the graph there is also a
loop-carried (anti-dependency) edge between $b$ and $a$. In the same way, for each (non-loop carried) anti-dependency edge
there exists a loop-carried data-flow edge in the other direction.

Because of this, typically each pair of nodes in the dependency graph is connected.
\footnote{An exception is the edge-case in which the dependency graph consists of two or more
unconnected subgraphs. In that case, the user can set
the maximum number of stages manually.} 
So, for each pair of operations we can compute the maximum distance they may be apart.
Call the greatest such distance  $d$. Now we know that in the modulo schedule, the first and last operation
may be at most $d$ cycles apart. So, the upper bound for the number of stages is $\ceil{\frac{d}{II}}$.

In more detail, we do the following:
\begin{enumerate}[label=\alph*.]
\item{Construct the graph $M$, with an edge from $b$ to $a$ with length $d \times II -l$ for each edge  
from $a$ to $b$ with latency $l$ and distance $d$ in the dependency graph.}
\item{Using the Floyd-Warshall\cite{FloydWarshall} algorithm for all-pair-shortest-paths, calculate the shortest path distance between
each pair of nodes in the graph. Note that the length of this path is the maximum distance that these nodes
may be apart.}
\item{Find the pair of nodes with greatest maximum distance they may be apart. Call this $d$.}
\item{The upper bound for the number of stages is now given by $\ceil{\frac{d}{II}}$;}
\end{enumerate}

At the end of \textbf{step 2}, we set $\MaxNumStages$ to the calculated upper bound.

Then, in \textbf{step 3},  
as described in the previous subsection, 
we create the SMT-problem for 
the current values of  $\II$ and $\NumStages$. 
In \textbf{step 4}, we ask the SMT-solver to solve the SMT-problem. This step is typically the most time-consuming
part of the algorithm.

There are two possible outcomes of \textbf{step 4}: the problem
is either satisfiable (SAT) or unsatisfiable (UNSAT)\footnote{
In our implementation, there is a third outcome: the user can provide a resource-limit to compiler, 
specifying the maximal amount of resources the compiler can spend per call to the SMT-solver before bailing out.
This can be used as a deterministic time-out functionality.  
}.
If the SMT-solver decides that the problem is satisfiable, then in \textbf{step 5} a schedule for the prolog,
epilog\footnote{As an optimization, in our implementation the prolog and epilog can (and are by default) 
scheduled in a separate step, after they are merged into the surrounding basic blocks.}, and kernel is generated from the satisfying assignment.
If the SMT-solver finds the SMT-problem to be unsatisfiable, then in \textbf{steps 6 and 7} the number of stages 
will be incremented  until a schedule is found or the upper bound for the number of stages is reached. 

When the number of stages has reached the upper bound without finding a schedule, 
the compiler will in \textbf{step 8} increment the $\II$, and start the search for a software pipelined schedule
of this $\II$ in \textbf{step 2}.

\subsection{Extending the algorithm with Register Pressure}
\label{sec:rp}

In the algorithm in Figure \ref{fig:algo}, we have not taken into account register allocation. 
This means that for loops with high register pressure, the algorithm may generate schedules for which no valid register-allocation
can be found. As we perform register-allocation after scheduling, the compiler has no (reliable) way of inserting spill-code and 
will exit with an error-message in that case.
In this subsection we explain how we deal with such loops.

One way to add support for loops with register pressure would be to add constraints for register-allocation to
the SMT-problem. That is, we could for each \emph{virtual register} in the loop, add a constraint 
that it should be mapped to one of the available physical registers in the processor. Next to that, we could
add constraints stating that when two virtual registers are mapped to the same physical register, their
live-ranges are not allowed to overlap.
We found, however, that this approach turned out to be too computationally expensive.

As an alternative for modelling register-allocation, we have decided to model \emph{register-pressure}.
That is, for each register-file, we model for each cycle, the number of virtual registers that is live
in that register file. We add the constraint that for each cycle, and register-file, the number of 
live registers is less than or equal to the register-file capacity.

Although this constraint is necessary for a valid register allocation to exist, it is not enough
to guarantee the existence of a valid register allocation. But in practice this works well:
In most cases constraining the register-pressure makes sure a valid register-allocation can be found. 
If no register-allocation can be found, the compiler can reschedule the code
using a tighter bound on register-pressure.

In the next subsection, we explain how the SMT-problem for SWP can be extended to deal with register-pressure.

In Figure \ref{fig:algo_rp}, we describe the algorithm for SWP with register-pressure. The two new steps
are \textbf{RP1} and \textbf{RP2}. In the algorithm, in \textbf{step 3} we create the original SMT-problem 
(that is, without taking into account register pressure). After a solution for this SMT-problem is
found, we check in \textbf{step RP1}, for this solution, for all register-files, whether the register-pressure is lower
or equal to the register-file size. If so, we are done. If, however, for some
register-files, the register-pressure exceeds the register-file capacity we go to \textbf{step RP2}.
In this step, for all the register-files for which the pressure exceeds the capacity, we add
the register-pressure constraints described in the next subsection to the SMT-problem
and will try to solve the SMT-problem again.

Note that we only add register-pressure constraints when register-pressure is detected in the solution
for the SMT-problem that does not consider register-pressure. Next to that, we only add constraints
for the subset of register-files for which pressure was detected. 
We use this \emph{lazy} approach to speed-up the compilation by 
avoiding the addition of unneeded constraints.

We can do so efficiently by making use of 
the \emph{incremental} feature of SMT-solvers: when, after the solver finds an SMT-problem 
to be satisfiable, we add additional constraints, the solver will not start from scratch:
instead, it will start from the original satisfying assignment and will reuse what it has
learnt about the problem thus far.

\subsection{Extension of the SMT Problem to deal with Register Pressure}

Below we describe the clauses and constraint we add to the SMT-problem to deal with register pressure:

We add the following variables:
\begin{enumerate}
\item{For each \emph{virtual register} $v$ and register-file $RF$, integer variables 
\begin{itemize}
\item{$\Start_{v}$ modeling the cycle in which $v$ starts becoming live.}
\item{$\End_{RF,v}$ modeling the cycle in which $v$ ends becoming live in register-file $RF$.}
\item{$\MinUse_{RF,v}$ modeling the cycle in which $v$ is used first in register-file $RF$.}
\item{$\MaxUse_{RF,v}$ modeling the cycle in which $v$ is used last in register-file $RF$.}
\item{$\MaxUseBeforeDef_{RF,v}$ modeling the cycle in which $v$ is used last before it is defined in register-file $RF$.}
\end{itemize}
}
\item{For each \emph{virtual register} $v$ and register-file $RF$, Boolean variables 
\begin{itemize}
\item{$\Used_{RF,v}$ modeling whether $v$ is used in register-file $RF$.}
\item{$\UsedBeforeDef_{RF,v}$ modeling whether $v$ is used before it is defined in register-file $RF$.}
\item{$\LiveOut_{RF,v}$ modeling whether $v$ is live-out in register-file $RF$.}
\end{itemize}
}

\item{For each \emph{virtual register} $v$, register-file $RF$, and cycle $c$, Boolean variables 
\begin{itemize}
\item{$\Live_{RF,v,c}$ modeling whether $v$ is live in register-file $RF$ at cycle $c$}
\end{itemize}
}
\end{enumerate}

The following constraints are added:

\begin{enumerate}

\item{Define $\Start$ in terms of $\Cycle$:}

For each virtual register $v$, let $d$ be the operation in the loop that defines $v$:

\[
\Start_{v} =
\begin{cases}
0 & \text{if no such operation exists} \\
\Cycle_{d} \% \II & \text{otherwise}\\
\end{cases}  
\]

\item{Define $\Used$, $\MinUse$, $MaxUse$, and $\MaxUseBeforeDef$ in terms of $\Cycle$ and $\Slot$}

For each virtual register $v$, let $U$ be the set of operations in the loop that uses $v$,
and for each register-file $RF$, 
let $S_{RF,u,v}$ be the set of issue-slots such that $v$ is read from register-file $RF$ if
$u$ is scheduled on the slot.  
\[
\Used_{v,RF} = \bigvee\limits_{\substack{u \in U \\ s \in S_{RF,u,v} }} \Slot_{u,s}
\]

\[
\MinUse_{v,RF} = \min(\{ \mathrm{if} \: \Slot_{u,s} \: \mathrm{then} \: \Cycle_u \% \II \: \mathrm{else} \: \II \: | \: u \in U , s \in S_{RF,u,v}\})
\]
\[
\MaxUse_{v,RF} = \max(\{ \mathrm{if} \: \Slot_{u,s} \: \mathrm{then} \: \Cycle_u \% \II \: \mathrm{else} \: -1 \: | \: u \in U , s \in S_{RF,u,v}\})
\]
\[
\MaxUseBeforeDef_{v,RF} = \max(\{ \mathrm{if} \: \Slot_{u,s} \land (\Cycle_u\%\II \leq \Start_{v}) \: \mathrm{then} \: \Cycle_u\%\II \: \mathrm{else} \: -1 \: | \: u \in U , s \in S_{RF,u,v}\})
\]

\item{Define $\UsedBeforeDef$ in terms of $\Used$, $\MinUse$ and $\Start$:}

For each virtual register $v$:

\[
\UsedBeforeDef_{v,RF} = \Used_{v,RF} \land \MinUse_{v,RF} \leq \Start_v 
\]

\item{Define $\End$ in terms of $\UsedBeforeDef$, $\MaxUse$, and $\MaxUseBeforeDef$.}

\[
\End_{v,RF} = \mathrm{if} \: \UsedBeforeDef_{v,RF} \: \mathrm{then} \: \MaxUseBeforeDef_{v,RF} \: \mathrm{else} \: \MaxUse_{v,RF} \:
\]

\item{Define $\LiveOut$ in terms of $\Slot$.}

For each virtual register $v$, let $U$ be the set of operations outside, but reachable from, the loop that use $v$,
and for each register-file $RF$, 
let $S_{RF,u,v}$ be the set of issue-slots such that operation $u$ reads virtual register $v$ from register-file $RF$ if
$u$ is scheduled on the slot.  

\[
\LiveOut_{v,RF} = \bigvee\limits_{\substack{u\in U \\ s \in S_{RF,u,v}}} \Slot_{u,s} 
\]

\item{Define $\Live$ in terms of $\Start$, $\End$, and $\UsedBeforeDef$.}

For each virtual register $v$, register-file $RF$,  and cycle $c$ such that $0 \leq c <  \II $:

\[
 \Live_{RF,v,c} = 
 \begin{cases} 
  c \leq \End_{RF,v} \lor \Start_{v} \leq c  &  \text{if } \UsedBeforeDef_{RF,v} \\ 
  \Start_{v} \leq c  \land (c \leq \End_{RF,v} \lor \LiveOut_{RF,v}) & \text{otherwise} \\
 \end{cases}
\]

\item{Register-pressure cannot be higher than the register capacity.}

For each register-file $RF$ with capacity $cap$ and cycle $c$ such that $0 \leq c <  \II $:
\[ \Count( \{ \Live_{RF,v,c} | v \in V \} ) \leq cap \]

Here $\Count$ is the function that counts the number of true expressions in a set.

\end{enumerate}

\section{Results}
\label{sec:results}

Our optimal software pipelining algorithm is used for scheduling more than 400 loops in firmware code for two generations of VLIW-processors used for wireless communication.
The largest of the two processors contains 8 issue-slots and 22 register-files. 

Although the algorithmic complexity of our approach is exponential, it is powerful enough to
generate schedules for all the tested loops.
The largest loop that was scheduled contains more than 250 operations.

We have evaluated the schedule quality of our approach on a set of 33 signal processing kernels.
Previously, these kernels were scheduled using a combination of heuristic algorithms and hand optimization.
For more than 80\% of the kernels our algorithm produced better results.
For the remaining 20\% of the kernels the results were equal. The maximum measured speed-up was $1.22$. The geometric average speed-up was $1.08$.

There are several advantages of our approach compared to the earlier used 
method of combining heuristic with hand optimization: 
\begin{enumerate}
\item{better schedules are generated,} 
\item{precious end-user time spent on tedious hand-optimization is saved,}
\item{the firmware engineer can be sure that an optimal schedule is generated, and}
\item{the user can ask the compiler for the \emph{reason} why a schedule with a certain initiation interval cannot be reached.}
\end{enumerate}

A disadvantage is that the invocation of the SMT-solver can increase the compilation-time to seconds, or for some large kernels, to a few minutes.

\section{Conclusion and Future Work}

\subsection{Conclusion}

Software pipelining (SWP) is a classic and important loop-optimization technique for VLIW-processors. 
It improves instruction-level parallelism with a limited increase in code size 
by compactly overlapping multiple iterations of a loop and executing them in parallel. 

Generating a Software Pipelined schedule for a loop is an NP-complete problem.
For this reason, SWP is typically implemented using heuristics that balance 
compile time and the quality of the generated schedule.  

However, as for VLIW processors, the schedule generated by the compiler directly determines the performance 
of the compiled application (for instance, scheduling an inner loop in 10 cycles
instead of 9, gives a performance degradation of that inner loop of a factor 0.9),
it makes sense to investigate optimal algorithms that are guaranteed to find a 
software pipelined loop with the smallest initiation interval. 

In this paper we have described an algorithm for optimal software pipelining
based on an SMT-solver. Although optimal approaches using Integer Programming (IP)
are known\cite{ExhSchedSurvey}, we are, as far as we know, 
the first to present an optimal software pipelining algorithm
using an SMT-solver.
 
We have shown that the algorithm is powerful enough
to schedule all inner loops of two generations of VLIW-processors for 
wireless communication.
We have also shown
that the schedule quality significantly outperforms the 
previously used combination of heuristic methods and hand optimization.

We have been the first to show how the \emph{unsatisfiable core} generation feature of modern SMT-solvers can be used by the compiler
to give feedback to programmers and processor-designers about \emph{why} a software pipelined schedule for a certain II is not possible.

\subsection{Future work}

A possible direction for future work is to speed-up the algorithm by exploiting
symmetries that appear in certain large loops. The idea is to prevent the solver 
from visiting potential solutions that are essentially equal to an earlier visited
solution because they are symmetric to each other. One way of doing so is by 
adding so-called symmetry breaking constraints\cite{Symmetry}. 

\section{Acknowledgements}

We would like to thank the Silicon Hive team, and particularly the compiler team, for providing the stimulating work environment that 
has made this work possible.

\section{References}

\bibliographystyle{abbrv}

\bibliography{Ref}

@inproceedings{Symmetry,
author = {Puget, Jean-Francois},
title = {On the Satisfiability of Symmetrical Constrained Satisfaction Problems},
year = {1993},
isbn = {3540568042},
publisher = {Springer-Verlag},
address = {Berlin, Heidelberg},
booktitle = {Proceedings of the 7th International Symposium on Methodologies for Intelligent Systems},
pages = {350–361},
numpages = {12},
series = {ISMIS '93}
}

@article{FloydWarshall,
author = {Floyd, Robert W.},
title = {Algorithm 97: Shortest path},
year = {1962},
issue_date = {June 1962},
publisher = {Association for Computing Machinery},
address = {New York, NY, USA},
volume = {5},
number = {6},
issn = {0001-0782},
url = {https://doi.org/10.1145/367766.368168},
doi = {10.1145/367766.368168},
journal = {Commun. ACM},
month = {jun},
pages = {345},
numpages = {1}
}

@book{VLIWBook,
author = {Fisher, Joseph and Faraboschi, Paolo and Young, Cliff},
year = {2005},
month = {01},
pages = {},
title = {Embedded computing: a VLIW approach to architecture, compilers and tools},
isbn = {978-1-55860-766-8}
}

@book{SMTBook,
author = {Kroening, Daniel and Strichman, Ofer},
title = {Decision Procedures: An Algorithmic Point of View},
year = {2008},
isbn = {3540741046},
publisher = {Springer Publishing Company, Incorporated},
edition = {1}
}

@inproceedings{SMS1,
author={Josep Llosa and  Antonio González and  Eduard Ayguadé and Mateo Valero},
title = {Swing Modulo Scheduling: A Lifetime-Sensitive Approach},
year = {1996},
publisher = {IEEE Computer Society},
address = {USA},
booktitle = {Proceedings of the 1996 Conference on Parallel Architectures and Compilation Techniques},
pages = {80},
series = {PACT '96}
}

@article{SMS2,
  author={Llosa, J. and Ayguade, E. and Gonzalez, A. and Valero, M. and Eckhardt, J.},
  journal={IEEE Transactions on Computers}, 
  title={Lifetime-sensitive modulo scheduling in a production environment}, 
  year={2001},
  volume={50},
  number={3},
  pages={234-249},
  doi={10.1109/12.910814}
}

@book{IPBook,
author = {Nemhauser, George L. and Wolsey, Laurence A.},
title = {Integer and combinatorial optimization},
year = {1988},
isbn = {047182819X},
publisher = {Wiley-Interscience},
address = {USA}
}

@article{ExhSchedSurvey,
author = {Lozano, Roberto Casta\~{n}eda and Schulte, Christian},
title = {Survey on Combinatorial Register Allocation and Instruction Scheduling},
year = {2019},
issue_date = {May 2020},
publisher = {Association for Computing Machinery},
address = {New York, NY, USA},
volume = {52},
number = {3},
issn = {0360-0300},
url = {https://doi.org/10.1145/3200920},
doi = {10.1145/3200920},
journal = {ACM Comput. Surv.},
month = {jun},
articleno = {62},
numpages = {50}
}

@article{IMSRau,
author = {Rau, B. Ramakrishna},
title = {Iterative Modulo Scheduling},
year = {1996},
issue_date = {February  1996},
publisher = {Kluwer Academic Publishers},
address = {USA},
volume = {24},
number = {1},
issn = {0885-7458},
url = {https://doi.org/10.1007/BF03356742},
doi = {10.1007/BF03356742},
journal = {Int. J. Parallel Program.},
month = {feb},
pages = {3–64},
numpages = {62}
}

@inproceedings{ModSchedRau,
  author = {Rau, B. Ramakrishna and Glaeser, Christopher D.},
  booktitle = {MICRO},
  crossref = {conf/micro/1981},
  editor = {Eckhouse, Dick},
  pages = {183-198},
  publisher = {IEEE/ACM},
  title = {Some scheduling techniques and an easily schedulable horizontal architecture for high performance scientific computing.},
  year = 1981
}

@proceedings{conf/micro/1981,
  editor       = {Dick Eckhouse},
  title        = {Proceedings of the 14th annual workshop on Microprogramming, {MICRO}
                  1981, Chatham (Cape Cod), Massachusetts, {USA}},
  publisher    = {{IEEE/ACM}},
  year         = {1981},
  url          = {http://dl.acm.org/citation.cfm?id=800075},
}

@inproceedings{practicalsat,
    author = {Joao Marques-silva},
    title = {Practical applications of boolean satisfiability},
    booktitle = {In Workshop on Discrete Event Systems (WODES},
    year = {2008},
    publisher = {IEEE Press}
}

@inproceedings{grasp,
 author = {J.P. Marques Silva and K.A. Sakallah},
 title = {GRASP - a new search algorithm for satisfiability},
 booktitle = {ICCAD '96: Proceedings of the 1996 IEEE/ACM international conference on Computer-aided design},
 year = {1996},
 isbn = {0-8186-7597-7},
 pages = {220--227},
 location = {San Jose, California, United States},
 publisher = {IEEE Computer Society},
 address = {Washington, DC, USA}
}

\end{document}